# In the Magma chamber: Update and challenges in ground-truth vulnerabilities revival for automatic input generator comparison

Timothée Riom, Sabine Houy, Bruno Kreyβig, Alexandre Bartel
*Umeå Universitet*
Umeå, Sweden
{firstname}.{lastname}@umu.se

*Abstract*—Fuzzing is a well-established technique for detecting bugs and vulnerabilities. With the surge of fuzzers and fuzzer platforms being developed such as AFL and OSSFuzz rises the necessity to benchmark these tools' performance. A common problem is that vulnerability benchmarks are based on bugs in old software releases. For this very reason, Magma introduced the notion of forward-porting to reintroduce vulnerable code in current software releases. While their results are promising, the state-of-the-art lacks an update on the maintainability of this approach over time. Indeed, adding the vulnerable code to a recent software version might either break its functionality or make the vulnerable code no longer reachable. We characterise the challenges with forward-porting by reassessing the portability of Magma's CVEs four years after its release and manually reintroducing the vulnerabilities in the current software versions. We find the straightforward process efficient for 17 of the 32 CVEs in our study. We further investigate why a trivial forward-porting process fails in the 15 other CVEs. This involves identifying the commits breaking the forward-porting process and reverting them in addition to the bug fix. While we manage to complete the process for nine of these CVEs, we provide an update on all 15 and explain the challenges we have been confronted with in this process. Thereby, we give the basis for future work towards a sustainable forward-ported fuzzing benchmark.

## I. INTRODUCTION

Fuzzing is now a prevalent technique to detect vulnerabilities after its first application in 1990, on testing Unix binaries' reliability [1]. Nowadays, software companies are continuously fuzzing their systems and applications. Google, for instance, fuzzes over a thousand different open-source projects through its OSS-Fuzz platform [2]. While AFL has become a reference, the number of fuzzing tools is constantly increasing. For instance, OSS-Fuzz relies on several of them, including libFuzzer [3], AFL++ [4], and Hongfuzz [5]. These fuzzers can have different exploration focuses, like code coverage [3], [5], [6], generating structured inputs [7], [8], respecting the specific language's syntax [9], [10], [11], [12], target specific components [13], [14], [15], focus on seed improvement [16], or specifically test certain vulnerability types in-depth, such as type confusion [17]. With this profusion arises the challenge of comparing fuzzers based on their performance.

Magma [18], a tool published in 2020, aims to tackle this challenge. It presents a method for assessing fuzzers' effectiveness in detecting real, publicly disclosed vulnerabilities in widely used C/C++ open-source projects. Magma's benchmark principle starts by reintroducing "*real* bugs in *real* software". The *real* bugs are former vulnerabilities from the very same project that are reintroduced in later versions by *reversing* the commit fixing the vulnerability. Hereby, Magma claims to enable a "*groundtruth fuzzing benchmark*". This holds ground on the base that the reversed patch prevents the exploitation of the vulnerability from this point forward: in code and in time. The resulting benchmark evaluates fuzzers' capacity to generate inputs that reach and trigger these reintroduced vulnerabilities. A task the article undertakes to also reflect on fuzzers' weakness-specific (i.e., vulnerability types) efficiency

This reintroduction of former vulnerabilities is called *forward-porting*. It is a manual task involving writing a patch that reverses the state of the files patched at their before-fix stage. Overall, the Magma benchmark covers nine open-source projects: libtiff [19], libpng [20], Lua [21], libxml2 [22], Poppler [23], OpenSSL [24], SQLite [25], libsndfile [26], and PHP [27]. It includes 112 registered CVEs [1] and six unreferenced bugs. Finally, the authors benchmark seven open-source fuzzers on vulnerable versions of these projects to measure and compare their ability to reach and trigger the implemented vulnerabilities.

This work explores how the *forward-porting* principle is maintainable on benchmarks over time. As the code of the target project evolves, reintroducing the vulnerability by reversing the patch becomes increasingly complex. Our motivation is to evaluate if specific cases arise where the reverse-patching breaks compilation, alters the capacity to trigger the vulnerability, or to reach it in exploitable conditions. We list these cases, categorise the changes that break the actual *forward-porting* of vulnerabilities and, overall, provide insights on Magma's benchmark principles four years after its release. We focus on maintaining the fairness and meaningfulness that Magma's *forward-porting* benchmark provides. Our concern is that specific changes introduced between a vulnerability patch and the latest release due to the usual software evolution, such as the addition of new functionalities or code refactoring, might actually affect the ability to *forward-port* CVEs and/or trigger forward-ported CVEs. As a result, fuzzers would then be evaluated on their capacity to complete unachievable tasks, for instance, if the *forward-ported* vulnerability cannot be reached. If this happens, the evaluation of fuzzers and their comparison would, over time, increasingly lose fairness and meaningfulness, respectively.

---

[1]CVE: Common Vulnerabilities and Exposures. A unique identifier for a specific vulnerability in given software versions. This catalog exists since September 1999

The state-of-the-art reference regarding bug-injection-based benchmark is FixReverter [28], both for the quality and quantity of inserted bugs. FixReverter detects three single-location-only fix patterns. These change patterns, consisting of conditional code structures, can be reverted to introduce a bug that FixReverter assumes to lie underneath. It facilitates the injection of around 8 000 bugs in 10 real-world programs. At this stage, the intrinsic reachability analysis has dropped 71% of candidate injection sites. Further, the fuzzing campaign that FixReverter carries on the base of five different fuzzers reached 24% of the injected bugs, and triggered 8% of them. In our work, we do not generate or use patterns from our set of CVEs. Our analysis detects and reverts vulnerability fixes and changes, which often spread across multiple functions and project files. We empirically evaluate and undertake the work necessary to keep the vulnerabilities both reachable and triggerable given the natural evolution of the target open-source software.

On a subset of Magma's CVEs (32), we update *Forward-Porting* based benchmarks through the following research questions:

*RQ 1.* **How functional is the *forward-porting* principle with the latest versions of target projects?** We aim to update the exploitability of vulnerabilities in the latest versions of target projects. In this research question, we "trivially" *forward-port* the once-deemed vulnerable portion of the code with a manual effort equivalent to Magma's patching. We find that a trivial reverse-patch of the fix fails to forward-port almost half of the 32 vulnerabilities.

*RQ 2.* **Is manual re-introduction of CVEs a sustainable strategy for maintaining *forward-porting*-based fuzzer benchmarks ?** We measure and report the effort necessary to make the modifications that both enable the building of a library and successfully port an exploitable vulnerability. Through extensive manual investigation, we were able to revive nine of the fifteen CVEs – for which trivial porting was not sufficient – in the latest version of the corresponding libraries. For the six others, we detail why our manual investigations were unsuccessful.

*RQ 3.* **What are the most usual changes that break the *forward-porting* of vulnerable code over time?** We analyze the different cases when porting a vulnerability either breaks the build of the target library or fails to trigger the vulnerable code. We observe that for roughly one in two cases, *checks and processing of the input* is the most likely type of change to break the *forward-porting*. It is followed by changes affecting the type or the variable's name for one in five cases.

Through these research questions, we intend to shed light on the challenges of *forward-porting* and help the community maintain a *forward-porting*-based benchmark. We highlight the effort required to update target projects to their latest versions, thereby ensuring the continued relevance of these benchmarks over time. We present our methodology in Section II, and associated results in Section III. We discuss the implications in Section IV. In Section V, we provide a literature perspective on the different techniques to evaluate and compare fuzzing tools. Finally, we conclude in Section VI.

## II. METHODOLOGY

We present the set of CVEs in Section II-A. Then, we iteratively describe the methodology used to tackle the three research questions in Sections II-B to II-D. Section II-E provides implementation and code availability details.

### A. Vulnerability Set

We rely on *32* vulnerabilities, listed in Table.I, from seven open-source projects: libpng (two CVEs), libTIFF (twelve CVEs), Lua (three CVEs), libxml2 (five CVEs), Poppler (four CVEs), php (three CVEs), and SQLite (three CVEs). Vulnerabilities in these projects are also used in Magma [18], a project containing the first dataset of *forward-ported* vulnerabilities to benchmark fuzzers. These vulnerabilities are selected based on our capacity to find publicly available Proof-of-Concepts (PoCs). For each of these PoCs, integration in our test set is preceded by the validation that these gathered PoCs both (1) trigger the vulnerable version, and (2) do not work in the fixed version. We validate the *forward-porting* on the basis that the specific PoC triggers the vulnerability. We admit that not triggering a vulnerability through one specific PoC is not a formal proof of exposure. Theoretically, a PoC variant could be crafted to suit the more recent context and trigger the vulnerability. However, if the vulnerability is triggered, it is empirical proof that the code is exposed and that it is fair to evaluate fuzzers on their capacity to trigger it. The absence of triggering also allows us to explain how changes in the target project may add protection layers on top of the fix.

One reason for not working on the entire Magma set, and to focus on a subset still representing one-fourth of the original, lies in the difficulty of finding publicly available and triggering PoCs. They are not found in the Magma repository, and if the artefacts contain 19 739 of fuzzer-generated crashing inputs unmapped to the CVEs nor to harness. Magma's authors write that they carried out a static analysis proving the reachability in triggering conditions in case of missing PoCs (they call Proof-of-Vulnerability). We rely on the efforts of the community for usable PoC public releases, which is a time- and skill-extensive task. The availability also depends on the release policy of the target libraries' owners. Such a transparent approach may put any instances that are not up-to-date at risk.

### B. RQ1. How functional is the *forward-porting* principle with the latest versions of target projects?

TABLE I: Dataset vulnerability types

| Heap-buffer-overflow | CVE-2013-7443, CVE-2016-10269, CVE-2016-10270, CVE-2018-8905, CVE-2016-1834, CVE-2016-1840, CVE-2018-18557, CVE-2019-9021, CVE-2019-9936, CVE-2019-11034, CVE-2019-11041, CVE-2019-12293 |
|---|---|
| NULL pointer dereferences | CVE-2018-7456, CVE-2019-7663, CVE-2019-10873, CVE-2019-14494, CVE-2019-19923, CVE-2020-24369 |
| Division-by-zeros | CVE-2016-10266, CVE-2016-10267, CVE-2018-13785 |
| Other types of out-of-bound writes | CVE-2015-8784, CVE-2016-5314, CVE-2017-9047 |
| Out of bound reads | CVE-2016-3658, CVE-2016-1762, CVE-2017-8872 |
| Memory exhaustion | CVE-2017-11613 |
| Stack-use-after-return | CVE-2019-7317 |
| Access of resource out of range | CVE-2020-15945 |
| Integer overflow | CVE-2019-9959 |
| Integer underflow | CVE-2020-24370 |

In this research question, we update the status of the *forward-portability* of the Magma benchmark's vulnerabilities w.r.t. latest version of target projects. Our evaluation features two steps: **(a)** assessing if trivial *forward-porting* is possible into the latest version of the target project; **(b)** verifying if the compiled library is vulnerable to the associated PoC.

We consider two approaches for step **(a)**. For the first one, we try to automate the reverse-patching of CVEs in latest versions, using `git` to generate patches. The second approach is more hands-on, as we manually revert the patched areas directly to their unpatched (i.e., vulnerable) version. Given Magma's authors manually wrote a patch for each CVE, suiting one specific version, and that we modify the files to their after-reverse-patching version: the work is comparable in type and load. Figure 1 illustrates the principle of *forward-porting* for CVE-2015-3784 from version v4.0.6 of libTIFF to v4.6.0. In Figure 1a, we observe the patch for the CVE with the lines removed from the vulnerable version in red, and the added lines highlighted in green. Figure 1b presents the latest version of the same lines of code. Finally, Figure 1c shows the *forward-ported* vulnerable code in v4.6.0.

In the second step, **(b)**, we execute the PoCs triggering the vulnerable versions on three different versions: a Reference version, an Intermediary version, and the latest version we consider in this work (usually in 2024). For the Reference, we settle on a first common git tag for all CVEs of a project that is -at best- the original *forward-porting* article [18] or closest to the fixing commits. The intermediary version is a release between the Reference and the latest version to highlight the temporal depth aspect. We finally reference the CVEs for which the PoC can trigger the latest *forward-ported* versions.

*C. RQ2.* Is the manual re-introduction of CVEs a sustainable strategy for maintaining *forward-porting*-based fuzzer benchmarks ?

This second research question focuses on PoCs that trigger the vulnerable version, yet fail to trigger the latest ones (from RQ1). Here, we investigate the causes of failures. Our

```
if( isTiled(tif) )
    imagewidth = tif->tif_dir.td_tilewidth;
+   tmsize_t op_offset = 0;
[...]
-       while (n-- > 0 && npixels <imagewidth)
+       while (n-- > 0 && npixels < imagewidth && op_offset < scanline)
            SETPIXEL(op, grey);
                if (npixels >= imagewidth)
                    break;
+       if (op_offset >= scanline ) {
+           TIFFErrorExt(tif->tif_clientdata, module, "Invalid data for
→ scanline %ld",
+               (long) tif->tif_row);
+           return (0);
+       }
            if (cc == 0)
                goto bad;
            n = *bp++; cc--;
```
(a) Fixing commit for CVE-2015-8784

```
    tmsize_t op_offset = 0;
    uint32_t imagewidth = tif->tif_dir.td_imagewidth;
    if (isTiled(tif))
        imagewidth = tif->tif_dir.td_tilewidth;
[...]
    while (n-- > 0 && npixels < imagewidth && op_offset < scanline)
        SETPIXEL(op, grey);
    if (npixels >= imagewidth)
        break;
    if (op_offset >= scanline)
    {
        TIFFErrorExtR(tif, module, "Invalid data for scanline %" PRIu32,
→ tif->tif_row);
        return (0);
    }
    if (cc == 0)
        goto bad;
    n = *bp++;
    cc--;
```
(b) Last version (v4.6.0) of libtiff/tif_next.c

```
//tmsize_t op_offset = 0;       //CVE-2015-8784
uint32_t imagewidth = tif->tif_dir.td_imagewidth;
if (isTiled(tif))
    imagewidth = tif->tif_dir.td_tilewidth;
    [...]
while (n-- > 0 && npixels < imagewidth)
    SETPIXEL(op, grey);
if (npixels >= imagewidth)
    break;
if (cc == 0)
    goto bad;
n = *bp++;
cc--;
```
(c) *Forward-Ported* code of CVE-2015-8784 in v4.6.0

Fig. 1: Re-introduction of the code from before fixing CVE-2015-8784 (i.e., from v4.0.6) in the latest context (i.e., v4.6.0).

objective is to evaluate the possibility of manually maintaining a dataset of vulnerabilities that can be triggered on the latest versions of a library. Failures in RQ1 can occur because the *forward-porting*: (i) breaks the target project's `build`, that the PoC itself is not compatible with the target project or that (ii) the PoC does not reach trigger conditions. To detect these breaking changes, we proceed through a dichotomy between releases. Our manual investigation compares variations in the flow of execution from the vulnerable version to the *forward-ported* ones. It involves the analysis of backtraces to follow the execution paths, the state of variables at different points in the code, and code logic changes. Alternatively, we sometimes use `git bisect` [29] to identify the commits behind the breaking changes: it is a considerably faster way to identify breaking commits, however, it fails to provide the analysis that would enable crafting a minimal *forward-porting* modification. Indeed, we observe that not all changes in a breaking commit may be mandatory to *forward-port* a CVE.

We consider the manual porting a failure in case the changes to carry are too considerable (number of lines to change),

too widespread (number of files to change), or include too many commits. Based on our experience, we set a limit of four commits beyond the fix in case of minimal changes. The limit relates to the issue of keeping track of the number of commits to revert sequentially in inverse order.

*D. RQ3.* What are the main reasons breaking the *forward-porting* of vulnerable code over time?

In this research question, we characterize the breaking commits identified in RQ2. We establish six different categories. A commit could introduce **variable name changes (C1)**. Adapting the code to the former or the fix to the new name is enough to port the vulnerability. Variables can also be affected by a more meaningful **change of type or structure (C2)**. This type of change can also be straightforward if it is not too widespread across files and versions. We also find some breaking changes related to a **removed functionality (C3)**, usually disabling the PoC's compatibility with the library. The **input (C4)** file can be better filtered, better **checked**, or better **processed** by the code. **Incorrect error handling (C5)** differs from the latter as they are inherent to inadequate treatment of errors already considered in the code. Finally, we also find non-functional **code refactorings (C6)** that prevent the *forward-porting*. If a commit breaks the *forward-porting* for several CVEs, we only count it once.

*E. Implementation*

One main contribution of our work is a reproducible methodology and a maintainable *forward-porting*-based benchmark. All experiments are, therefore, docker-contained. Each container contains a set of tools to *checkout* to the target version of the library, apply *forward-porting* modifications to either execute the related PoC, and/or to *forward-port* a vulnerability.

In general, PoC execution successes are detected through LLVM's or GCC's Address Sanitizer or Valgrind. LibTIFF's CVE-2017-11613 stands out as a memory exhaustion we detect when the program hangs.

All code and data can be anonymously accessed by reviewers online at https://osf.io/xnejq/?view_only= 9d23901bf9d04cd4968c70e950d4f95f.

### III. RESULTS

*A. RQ1.* How functional is the *forward-porting* principle with the latest versions of target projects?

Step **(a)**: We evaluate the status of trivial *forward-porting* unpatched code chunks into latest versions of the target projects. In other words, we evaluate the longevity of trivial *forward-porting* operations. An automated `git` based approach (*i.e.,* creating a patch and applying it in reverse on the later versions) does not work directly on the latest version for any of the target projects. This implies that, in all cases, the target library's source code evolved enough over time to prevent this simple automated approach based on `git`. These observations justify the need for another, more *hands-on*, approach. Therefore, we rewrite each file patched in the fix for each version (ref., inter., and latest), vulnerability, and target library. As a result, the area of the exposure (i.e., the area fixed in the patch) looks like the same code before fixing.

Step **(b)**: Results for RQ.1 are shown in Table II. Specifically, the results on the latest versions are presented in the last two lines. A tick (✓) indicates that the PoC successfully triggered the *forward-ported* vulnerability. A cross (✗) indicates that the PoC did not trigger the *forward-ported* vulnerability. We investigate the root causes of these failures in RQ.2.

In the Reference versions of the projects, 24 CVEs are triggered after *forward-porting*.

**Success:** For the latest version, the **trivial** *forward-porting* works directly in 17 cases out of the 32 vulnerabilities analysed. It is the case for both of libpng's CVEs, one CVE for Lua, five CVEs for libTIFF, one for Poppler, and two for SQLite. As for libxml2, the trivial *forward-porting* reaches the latest version available (v2.12.0) for two CVEs.

**Failure:** For the 15 other vulnerabilities, it is **not possible to trigger** the re-injected vulnerabilities with the PoC.

- It is either because the trivial *forward-porting* of the once-vulnerable to the latest version of the target code **breaks** the build of the target library. Hence, this vulnerable version of the code is no longer compatible with the latest version of the target library. This is the case for CVE-2020-15945 (Lua), CVE-2020-24370, and CVE-2016-3658 (libTIFF).
- Or, in case the build passes, launching the PoC on the build library fails to trigger the vulnerability. Two behaviours, further described in RQ2, can be observed: Either the PoC fails to launch (i.e., the PoC is no longer compatible with the project), or the execution of the PoC finishes without any error nor triggering of the catching method (ASan, Valgrind, ...). For two of libTIFF's CVEs the PoC fails to launch (CVE-2016-5314 and CVE-2016-10267). In the remaining cases, the program terminates without triggering the vulnerability.

> Trivial *forward-porting* loses its capacity to *revive* the vulnerabilities over time. The success rate drops from 75% in the reference version to 53% in the latest versions. This confirms the expectation that, with time, more work will be required to maintain the vulnerabilities in the benchmark.

*B. RQ2.* Is manual re-introduction of CVEs a sustainable strategy for maintaining *forward-porting*-based fuzzer benchmarks ?

In this research question, we detect breaking commits and fix the issues preventing the *forward-portability* of old CVEs into later versions of the target projects. To investigate the root causes of the failures, we manually analyse the 15 vulnerabilities from Section III-A for which the forward-porting was unsuccessful. The results are provided in Table IV, and the commits are listed in Table V (see in Section III-C). Ticks (✓) represent cases for which we manage to *port* the

TABLE II: RQ1: Execution of PoCs different *forward-porting* versions

| | libpng | | Lua | | | libTIFF | | | | | | | | | | |
|---|---|---|---|---|---|---|---|---|---|---|---|---|---|---|---|---|
| | CVE-2018-13785 | CVE-2019-7317 | CVE-2020-15945 | CVE-2020-24369 | CVE-2020-24370 | CVE-2015-8784 | CVE-2016-3658 | CVE-2016-5314 | CVE-2016-10266 | CVE-2016-10267 | CVE-2016-10269 | CVE-2016-10270 | CVE-2017-11613 | CVE-2018-8665 | CVE-2018-7456 | CVE-2018-18557 | CVE-2019-7663 |
| Reference version | v1.6.38 | | v5.4.1 | | | v4.1.0 | | | | | | | | | | | |
| Triggers | ✓ | ✓ | ✓ | ✓ | ✓ | ✓ | ✗ | ✗ | ✗ | ✓ | ✓ | ✗ | ✓ | ✓ | ✓ | ✓ | ✗ |
| Intermediary ver. | 1.6.39 | | v5.4.4 | | | v4.3.0 | | | | | | | | | | | |
| Triggers | ✓ | ✓ | ✗ | ✓ | ✗ | ✓ | ✗ | ✗ | ✗ | ✗ | ✓ | ✗ | ✓ | ✗ | ✓ | ✓ | ✗ |
| Latest version | v1.6.40 | | v5.4.7 | | | v4.6.0 | | | | | | | | | | | |
| Triggers | ✓ | ✓ | ✗ | ✓ | ✗ | ✓ | ✗ | ✗ | ✗ | ✗ | ✓ | ✗ | ✓ | ✗ | ✓ | ✓ | ✗ |

| | libxml2 | | | | | Poppler | | | | Php | | | SQLite | | |
|---|---|---|---|---|---|---|---|---|---|---|---|---|---|---|---|
| | CVE-2016-1762 | CVE-2016-1834 | CVE-2016-1840 | CVE-2017-8872 | CVE-2017-9047 | CVE-2019-9959 | CVE-2019-10873 | CVE-2019-12293 | CVE-2019-14494 | CVE-2019-9021 | CVE-2019-11034 | CVE-2019-11039 | CVE-2013-7443 | CVE-2019-9936 | CVE-2019-19923 |
| Reference version | v2.9.10 | | | | | 21.08.0 | | | | 7.4.0 | | | fix [1] | v3.30.0 | |
| Triggers | ✗ | ✓ | ✗ | ✓ | ✓ | ✓ | ✓ | ✓ | ✓ | ✓ | - | ✓ | ✓ | ✓ | ✓ |
| Intermediary ver. | 2.10.0 | | | | | 23.01.0 | | | | 8.0.0 | | | 3.35.0 | | |
| Triggers | ✗ | ✓ | ✗ | ✗ | ✓ | ✗ | ✓ | ✓ | ✓ | ✓ | ✓ | ✓ | ✗ | ✓ | ✓ |
| Latest version | v2.12.0 | | | | | 24.09.0 | | | | 8.3.12 | | | 3.47.0 | | |
| Triggers | ✗ | ✓ | ✗ | ✗ | ✓ | ✗ | ✓ | ✓ | ✓ | ✓ | ✓ | ✓ | ✗ | ✓ | ✗ |

[1]:version-3.8.2

TABLE III: RQ2: Status of the PoC executions on CVEs with reverting of breaking changes

| | Lua | | libTIFF | | | | | | | libxml2 | | | Poppler | SQLite | |
|---|---|---|---|---|---|---|---|---|---|---|---|---|---|---|---|
| | CVE-2020-15945 | CVE-2020-24370 | CVE-2016-3658 | CVE-2016-5314 | CVE-2016-10266 | CVE-2016-10267 | CVE-2016-10270 | CVE-2018-8905 | CVE-2019-7663 | CVE-2016-1762 | CVE-2016-1840 | CVE-2017-8872 | CVE-2019-9959 | CVE-2013-7443 | CVE-2019-19923 |
| Reference version | v5.4.1 | | v4.1.0 | | | | | | | v2.9.10 | | | 21.08.0 | v3.30.0 | |
| Triggers | | | ✓ | ✓ | ✓ | ✓ | ✗ | ✓ | ✓ | ✗ | ✓ | | ✓ | ✓ | |
| Intermediary vers. | v5.4.4 | | v4.3.0 | | | | | | | 2.10.0 | | | 23.01.0 | 3.35.0 | |
| Triggers | ✓ | ✓ | ✓ | ✓ | ✓ | ✓ | ✗ | ✓ | ✓ | ✗ | ✓ | ✓ | ✗ | ✓ | |
| Latest version | v5.4.7 | | v4.6.0 | | | | | | | v2.12.0 | | | 24.09.0 | 3.47.0 | |
| Triggers | ✓ | ✓ | ✓ | ✗ | ✓ | ✗ | ✗ | ✓ | ✓ | ✗ | ✓ | ✗ | ✗ | ✓ | ✓ |
| # reversed commits | 3 | 1 | 1 | 4 | 1 | 4 | 3 | 2 | 3 | 3 | 1 | 3 | 4 | 1 | 3 |

exposure of target projects to original PoCs. Our investigation highlights that there usually is more than one PoC-preventing commit, i.e., bug fixing change, per CVE.

*1) Successful forward-porting:*

**Lua:** CVE-2020-15945 requires tracking evolutions of the instruction pointer (`oldpc`) implementation in the Lua language. Triggering the vulnerability in v5.4.4 can be done through a partial reverse-patching of the commit `949187b0`. To further trigger the segmentation fault in v5.4.6, we must cherrypick a few changes to revert from `9b4f39a`. For instance, reverting only to v5.4.4's version of the methods `getshstr` in `ldebug.c` and `getlngstr` in `lvm.c` suffices. The commit message of this last commit states that it prepares for string input sanitisation to prepare for forthcoming opening to the newest input types. CVE-2020-24370 is even simpler as a variable type change from `*ci->func` to `ci->func.p` enables to pass the build and the exposure.

**libTIFF:** We can also trigger CVE-2016-3658 for all versions tested by only renaming the variable `tif_dir-.td_stripoffset` to `tif_dir.td_stripoffset_p`. This suffix alone was enough to break the trivial *forward-porting* for this CVE. Re-enabling CVE-2016-10266 in the latest versions requires, at first, minimally invasive changes. However, the analysis to detect the first breaking change is less straightforward. It is possible through the understanding of libTIFF's copy tool (`tiffcp`). Opening input files starts with evaluating their encoding (in strips, tiles, or rows). This evaluation evolves in `9e9a0bbf` to prevent overreading image source data by switching the encoded-based copy function. The new computation switches which encoded-based copy function (`cf`) is selected and prevents the exposed path in `cpDecodedStrips` to be executed. It further takes to respect C99 type standards (commit `39a74eed`) and to split according to CVE-2016-10270's fix's refactoring. *Forward-porting* CVE-2018-8905 first requires undoing commit `f13cf46b` to trigger v4.2.0. As such, we re-enable the size of one-strip-clamped input to be misinterpreted. Then, we can easily re-enable the `-ignore-errors` (or `-i`) input option for the copy tool `tiffcp` to port up to v4.6.0, included. CVE-2019-7663 can be reintroduced in v4.1.0 and v4.3.0 by disabling the input checks introduced by two commits. It also requires to re-enable the `-i` option mentioned above.

**Libxml2:** CVE-2016-1840 is reintroduced to the latest version by reverting commit `fb56f80e` on top of the fix.

**SQLite:** Triggering CVE-2013-7443 only requires identifying one commit related to increasing the default size of objects on the heap. To port CVE-2019-19923 to the latest version, three commits have to be reverted from 3.41.0 to 3.47.0. The first two are related to the treatment of left outer-joins and the last one to iterate based on specific boundary-set blocks of data in memory instead of reading them as a stream as they come.

*2) Incomplete manual forward-porting:*

**LibTIFF:** CVE-2016-5314 and CVE-2016-10267 can be forward-ported successfully until v4.5.1 when the tools, resp., `rgb2ycbcr` and `tiffmedian` are removed – just 27 commits before `v4.6.0`. Each of the four breaking commits needs to be identified and reverted to port the CVEs to this point. In the case of CVE-2016-5314, the tool employed by the PoC is removed for the first time even before the vulnerability is fixed. Since then, the source code has been still present but no longer compiled. CVE-2016-10270 stops being manually maintainable by the rules set in Section II-C after the third breaking commit, which happens before even reaching v4.1.0. This commit modifies 18 files, including above 40 code chunks in `tif_dirread.c` – the main file that the fix modifies.

**Libxml2:** CVE-2016-1762 affects v2.9.3. The trivial *forward-porting*, however, fails before v2.9.4. It is necessary to revert commit `0bcd05c5` also on `parserInternals.c` to reach this tag. This commit sanitises non-UTF-8 charsets, preventing the library from missing the end of input. Further attempting to reach 2.9.5 requires to revert `5f440d8c` and also `aa267cd1`. The task becomes increasingly tedious, with already four commits to revert while only covering 163 of the 1 759 commits on the way to v2.12.0.
Neither did we manage to carry CVE-2017-8872 to the latest v2.12.0. We can find two first breaking commits (ASCII filtering and buffer allocation computation) before reaching `facc2a0`. This commit includes 30+ chunks and makes a new manual dichotomy round time-consuming, given the four other commits to revert.

**Poppler:** CVE-2019-9959's list of commits to revert also increases fast in early versions: before reaching version 22.04.0, it is already three commits to revert plus the patch. The dynamic analysis of the call-stack pinpoints commit `a6b2442e` to be responsible for breaking the *forward-porting*. This commit is spread out over fourteen files, substituting various variable types for safer ones. As such, the type `std::vector<unsigned char>` replaces `unsigned char` and allows for tighter memory management. The research for a manually crafted patch is laborious from there on, as the level of modification to carry is now significant, and is likely to increase before reaching v24.09.0.

> Through manual modifications, such as changing variable types, copying/pasting previous versions of some code chunks, or removing conditions, we completed the *forward-porting* to the latest version for 9 out of 15 CVEs. Together with the trivial *forward-porting* (cf RQ.1), the manual maintenance of forward-ported vulnerabilities is realistic for 76% of cases after 4 years. However, as the target library continues to evolve, the process is likely to become increasingly complex. In the remaining six cases, we had to abort manual *forward-porting* since the number of commits or files to change became too large to complete and carry further along with the libraries' regular evolution in time.

*C. RQ3.* What are the main reasons breaking the *forward-porting* of vulnerable code over time?

We present, in Figure 2, the temporality for each breaking commit for which the trivial *forward-porting* fails (RQ1). Each project graph spans from the earliest fix of its CVEs to the latest version considered. The overall activity of each project is displayed in the background (clearest colour) through the two-week number of commits. Every CVE gets a lifeline from fix to latest, on which a triangle highlights the breaking commits, and vertical bars represent a specific CVE-related activity. This activity is counted in the number of commits that affect the files modified by either the fix or the breaking commits –in every two weeks. The project- and CVE-specific activities use different scales, which enables the display of all CVEs of one library in one figure. The scale is, however, the same between CVEs of the same project.

The figures highlight the relationship between high activity on the project and the introduction of breaking commits. For instance, the breaking commits of Lua's CVE-2020-24370, the second one of libTIFF's CVE-2016-10267, or libxml2's first and last breaking commits. Also, Poppler's late 2021 code burst led to the second breaking commit. This figure also highlights the distance to the latest version at which we reach the conditions to stop the manual investigation. The last breaking commit we consider for CVE-2016-10270 occurs before version 4.1.0 and even before the fourth code burst. Overall, the activity on SQLite is very high, making it difficult to draw conclusions based on this, but the last breaking commits of CVE-2019-19923 happened during a higher overall activity.

We categorise the breaking events into six categories that we present in Table IV. Symbols in the left maps for Table V's breaking commits' categorisation. If a commit is not attributed a symbol from Table IV it is because this commit is already a breaking commit for another CVE of the project. The last column provides whether the PoC triggers the CVE is the latest version attempted.

TABLE IV: Categorisation of commits breaking the *forward-porting*

| | | | |
|---|---|---|---|
| □ | C1 | Variable name change | 1 |
| △ | C2 | Variable change of type or structure | 5 |
| ▽ | C3 | Removed functionality | 3 |
| ● | C4 | Input check and processing | 18 |
| ε | C5 | Incorrect error handling | 3 |
| o | C6 | Code refactoring | 3 |

TABLE V: List and categorisation of commits breaking *forward-porting* of each CVE

| **Lua** | | PoC |
|---|---|---|
| CVE-2020-15945 | 949187b0● 9b4f39a● | ✓ |
| CVE-2020-24370 | 413a393e△ | ✓ |
| **libTIFF** | | |
| CVE-2016-3658 | 371ad265□ | ✓ |
| CVE-2016-5314 | 30366c9f▽ ec4d8e08● 56a1976eε eab89a62▽ | ✗ |
| CVE-2016-10266 | 9e9a0bbf● | ✓ |
| CVE-2016-10267 | 2e822691● 39a74eed△ 072cbbebε eab89a62 | ✗ |
| CVE-2016-10270 | 7057734d● 0489f1f8o 371ad265 | ✗ |
| CVE-2018-8905 | f13cf46b● 280a568a▽ | ✓ |
| CVE-2019-7663 | 2b0d0e69● 7b1f03c3● 280a568a | ✗ |
| **libxml2** | | |
| CVE-2016-1762 | 0bcd05c5● 5f440d8c● aa267cd1● | ✗ |
| CVE-2016-1840 | fb56f80e● | ✓ |
| CVE-2017-8872 | 4fd69f3● cabde70● facc2a0ε | ✗ |
| **poppler** | | |
| CVE-2019-9959 | 814fbda28o 7d7e09cfo 541e777△ a6b2442△ | ✗ |
| **sqlite** | | |
| CVE-2013-7443 | 56d65cd7b9△ | ✓ |
| CVE-2019-19923 | d198183465● ee37302095● 3c8e438583● | ✓ |

**Variable evolution:** For seven CVEs, we find an evolution of variables to cause the `build` to break. It can be a simple change of the variable name or a more consistent change of its structure. For CVE-2016-3658, we find a patch appending a simple `_p` suffix, standing for "protected". This change alone breaks the *forward-porting* of the fix. In six other cases, the variable type and subsequent adaptation suffice. 2020-15945's fix is also counted as a breaking commit because the change of the variable structure has side effects on the project. For instance, a field of the new object holds the variable that was previously used. Regarding CVE-2020-24370's `413a393e`, a simple variable type change from `*ci->func` to `ci->func.p` enables to pass the build and trigger the exposure. The *forward-porting* of CVE-2019-9959 stops to be efficient at `a6b244c`, when objects structured in unsigned char are replaced by `std::vector<unsigned char>`. The latter type provides tools to control, clean, and limit the size of elements, preventing the exploitation. We have another example in SQLite's CVE-2013-7443: heap allocation of objects is restructured, including an increase of their maximal size (`56d65cd7b9`) preventing the original PoC payload.

**Removed functionality:** The evolution of the target library can also change available features and/or generated binaries to run the PoC. For example, in libTIFF, `tiffmedian` is removed just before the latest version (v4.6.0) for CVE-2016-10267. Also, the tool used for CVE-2016-5314's PoC (`rgb2ycbcr`) is also removed even before the vulnerability is patched. The commits affect the `Makefiles` that stopped building these binaries before they were properly archived and removed. Regarding libTIFF's PoCs using `tiffcp`, version v4.6.0 removes the `-i` option that enables treating files without a complete check of the input file structure.

**Input check and processing:** Another issue arises when newer input checks add a second layer of protection on top of the fix. As an example, CVE-2016-1840 further requires to revert commit `fb56f80e`, introducing assessment of the position of the end of input character before reading. The original fixing commit circumvents the original issue (i.e., a quick fix). CVE-2017-8872 is further disabled by properly checking invalid HTML tags and better computing the available space in relation to the input. CVE-2019-7663 also falls in this category with commits `2b0d0e69` and `7cc76e9b` checking the shape of the image file before treating the data. Regarding CVE-2016-10266's `9e9a0bbf` limits the space to allocate instead of letting the file claim it for a figure. It is worth noting that this commit is also patching CVE-2016-10270, another vulnerability in libTIFF. CVE-2016-1762's commits are all related to parsing the input: determining the end of the input, escaping characters, and boundary checks. SQLite's CVE-2019-19923 trivial *forward-porting* breaks at `d198183465` which improves filtering queries on left outer joins.

**Code refactoring:** Regarding libTIFF's `0489f1f8`: one chunk is split between different functions. This prevents trivial un-patching as we also need to split the reverse-patching to reintroduce the vulnerability. CVE-2019-9959's (Poppler) two first breaking commits adapt the code to the C99 syntax standards and remove conditions deemed unnecessary by then.

**Incorrect error handling:** This category contains libTIFF's `56a1976e` that was inflating the input while an error was already detected. `072cbbeb` was also assuming the whole block (or strile) of the input was read at a point when it should assess for an error status. Finally, libxml2's `facc2a0` protects the EOF status from modification when the function should return.

> While breaking commits' introduction often occurs with high activity on fix-related files, some appear at calmer times, which does not help with their prediction. Among them, additional input checks are the most frequent reason to break the trivial *forward-porting*(≈53%). The second type of change breaking the capacity to revive a vulnerability is either affecting the variables' name, type, or structure (≈21%).

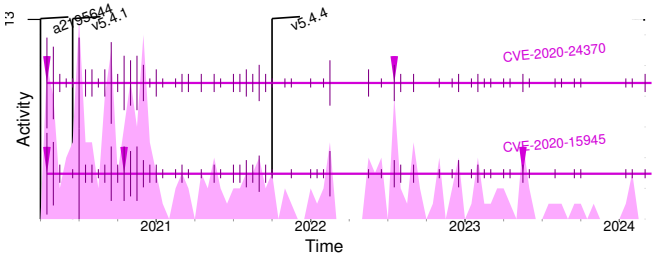

(a) Breaking commits and activity of Lua CVEs

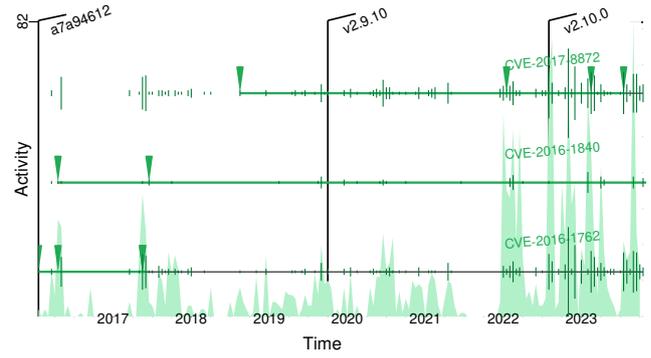

(c) Breaking commits and activity of libxml2 CVEs

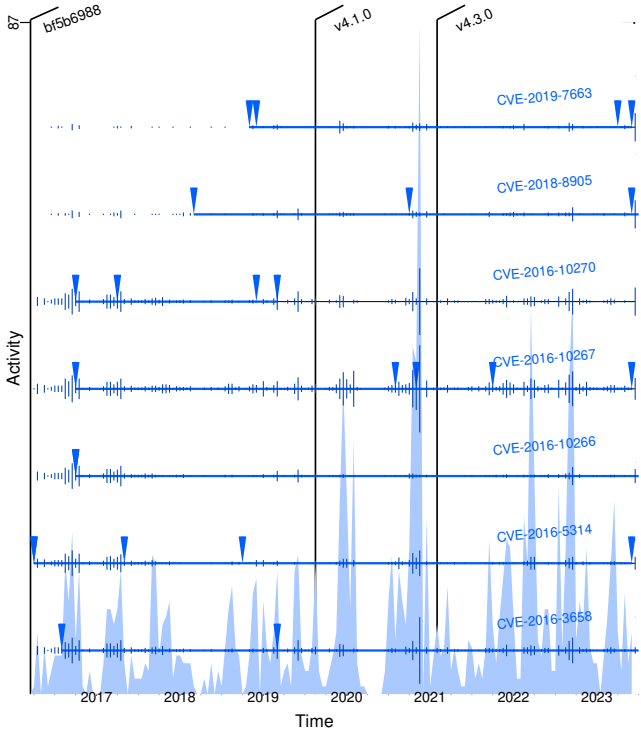

(b) Breaking commits and activity of libTIFF CVEs

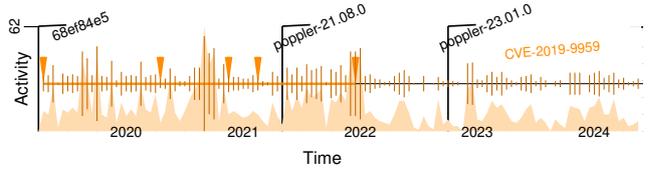

(d) Breaking commits and activity of Poppler CVEs

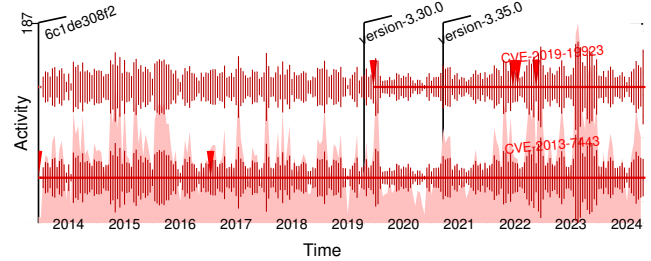

(e) Breaking commits and activity of SQLite CVEs

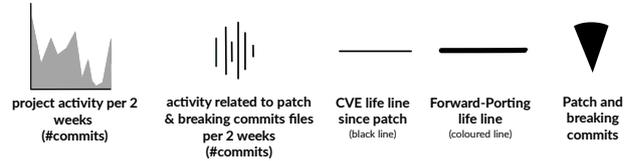

(f) Legend

Fig. 2: Chronologies for breaking commits projects and CVE-related activities in Lua, libTIFF, libxml2, Poppler and SQLite

## IV. DISCUSSION

### A. Limitations

The main limitation of our approach revolves around our evaluation the presence of a CVE through its exploitation by one specific PoC. We more precisely revive one exploitation path. Despite the presence of the fix and breaking commits, that we identify in this work, other paths may exist. Evaluating the presence of such alternative paths could benefit from a symbolic execution analysis, collecting all conditions to reach vulnerable areas of the code and comparing them to the exploitation conditions. Such work would provide PoC variants; evaluating CVE exploitability more accurately.

Another limitation regards the `git-bisect` approach, forthrightly removing entire commits. It is conceivable that only parts of the breaking commits are truly necessary to *forward-port* a CVE. A work evaluating partial commit reverting would help reduce the process's side effects on the rest of the targeted open-source software. It could further help reduce the number of breaking commits.

Resolving these limitations would improve the overall quality of the *forward-porting* process. However, our approach provides definitive evidence that the CVE is brought back in exploitable conditions. A key issue in bug coverage-based benchmarks such as Magma's.

### B. Challenges

While throughout the study, we encountered no condition making the forward-porting process categorically impossible, in some cases, we had to abort because the manual effort was no longer sustainable. This is the case if changes are spread out across too many files or code blocks, or the sequence of commits to revert grows too large.

For example, in CVE-2019-9959, after finding the fourth commit to revert, we need to revert 14 files to revive the vulnerability. From this point onward, we would have to revert the changes in these 14 files to each attempt-commit. More precisely, it takes us 8-13 attempts to find the next breaking commit with the dichotomy. For each of these attempts, we must revert all the changes in the 14 files again. Without automation, this is impractical. The same applies if the number of lines to change within a file grows too large (e.g., CVE-2017-8872 with 30 chunks).

We also stop if the number of commits to revert grows larger than four. This threshold is chosen empirically based on the manual effort we encountered in such cases. With each additional commit to revert and when working through the dichotomy, at each attempt, we have to revert the entire chain of previously found breaking commits one by one. That is because an earlier commit may only be reversible with the changes of a later commit.

From these challenges (size and number of commits), we conclude that a **manually** crafted forward-ported benchmark is possible but becomes hardly maintainable over time. Additionally, manual forward-porting can require domain-specific knowledge and functional understanding of the libraries, such as understanding libTIFF's `tiffcp` tool.

### C. Maintanability of a Forward-Porting based Fuzzer Benchmark

Beyond the challenges of *forward-porting* individual CVEs, we propose three exclusion rules helping to maintain such a benchmark up to date. The **complexity** to carry further the manual forward-porting, mentioned just above, stands as our first exclusion rule **(1)**. Seven CVEs out of the thirty-two we analysed would need to be excluded based on this criterion. Another aspect regards the **intercompatibility** of CVEs altogether. The target of a benchmark is not to launch a fuzzer campaign on different versions of the same target library, each including only one *forward-ported* CVE. This approach would be highly inefficient compared to providing one up-to-date version of each target library containing all CVEs. However, reversing specific breaking commits to make one CVE exploitable may prevent another vulnerability from being altogether vulnerable. For instance, Lua's CVE-2020-24369 reverse fixing in `changedline` of `ldebug.c` is incompatible with our rewriting of CVE-2020-15945 into v5.4.6. For this exclusion rule **(2)**, there can be two solutions to adopt: either **a)** we keep the latest CVEs (considering they are more meaningful to keep) or **b)** keep the biggest set of inter-compatible CVEs. For instance, if we drop libxml2's CVE-2016-1762, all four other *forward-ported* CVEs are compatible in the last commit (`facc2a0`) to respect the complexity rule for the four CVEs. We determine another exclusion rule **(3)** regarding the fallout of the *forward-porting* on the **functionality** of the resulting library. We can use the test suite included in the projects to assess such aspects. The "`make test`" option for libpng with both CVEs ported (and not conflicting): 7 out of 32 tests fail - related to the reverse-patching for CVE-2018-13785 to pass. The v1.6.40 of Lua's CVE-2020-24369 also fails the `./all` testsuite on one test (`db.all`) while our maintained CVE-2020-24370 passes them all (30+). However, our maintained version of CVE-2020-15945 breaks most of them. An assessment of the tests shall determine which test we should allow failure for the CVE to be kept. The first example is anti-fallback tests, added alongside the CVE fix.

These criteria, targeting to maintain the quality of such a benchmark, would, however, have a detrimental impact on the quantity of CVEs that are considered. Hence, the pool of CVEs also has to be increased for each benchmark revision. Listing all new CVEs from the NIST's feed [30] provides 27 new candidates for libxml2 and 68 candidates for libTIFF since 2021. Eventually, a yearly revised benchmark could stand as a satisfying solution.

### D. Evaluation of the workload

The manual forward-porting took over one year to complete. It involved different analysis techniques for identifying breaking commits which makes it difficult to give a precise and weighted measurement of the per-CVE manual effort. Our increasing experience with the process improved the tools and the methodology. The most in-depth analysis could require a week of combined reverse engineering techniques so as to identify one breaking commit, while the `git-bisect` approach was faster yet blunt.

### E. Automation of the forward-porting process

Based on the previous remarks and our experience in identifying commits that break *forward-porting*, we plan to work on automating the *forward-porting* process. We will consider different granularity levels for the reverse patching automation, such as: all files authorised, patch-files only, function level or chunk level. The tighter the granularity, the more optimised and the smaller the *forward-porting* patch from fix to the latest version.

One benefit of optimisation is to reduce *forward-porting* side-effects on all three exclusion rules cited above: less code to reverse-patch across versions (**complexity**), less interaction from one CVE with another (**intercompatibility**), and with the project's **functionality**.

## V. RELATED WORK

### A. Fuzzing

In "Fuzzing: State of the Art", Liang et al. [31] discuss the general process and classifications of fuzzing, as well as key obstacles and state-of-the-art technologies used to overcome them. Authors investigate and classify several widely used fuzzing tools, including AFL [6], Peach Fuzzer [32], and libFuzzer [3]. Their primary goal is to provide a better understanding of fuzzing and potential solutions for improving fuzzing methods in software testing and security. One of the research questions addressed in the article is the future directions for fuzzing. The authors discuss common problems, such as path explosion, oracle, coverage, and efficiency problems, and provide suggestions to answer them.

*B. Fuzzer Evaluation*

The article "Evaluating Fuzz Testing" by Klees et al. [33] investigates the experimental evaluations of fuzz testing techniques, strategies, and algorithms used to discover security-critical bugs in real software. The authors surveyed 32 fuzzing articles and found problems in every evaluation they considered. They further experimentally showed that these evaluation problems translate to wrong or misleading assessments. Unifuzz [34] started to address the issue by comparing 35 different fuzzers (eight presented in the article) for real-world bug detection over 20 real-world programs. The authors provide some solid guidelines to consider when evaluating fuzzers. Our research could potentially complement their findings on fuzz testing evaluation by investigating the use of an approach like *forward-porting* when doing a fuzz evaluation. In "FuzzBench: A Free Service for Evaluating Fuzzers" [35], researchers introduce FuzzBench, an alternative platform for fuzzer evaluation. The authors present case studies demonstrating the effectiveness of FuzzBench in evaluating various fuzzers, including AFL, Honggfuzz, and LibFuzzer. They show that FuzzBench can uncover bugs missed by other evaluation methods and help identify the strengths and weaknesses of different fuzzers. The authors draw a few comparisons to Magma 's [18] benchmark throughout the paper. They note that Magma's *forward-porting* approach of evaluating fuzzers is solely based on bug coverage and can be misleading due to the sparse distribution of real-world bugs in programs. They advocate for a combined evaluation metric including code and bug coverage to provide a more comprehensive evaluation of fuzzers.

*C. Bug introduction*

To our knowledge *Magma* describes the only methodology for reintroducing real-world bugs. Synthetic approaches can be found in [36], [37], [28], [38]. Fixreverter [28] is a pattern-based tool to locate syntactic and semantic patterns of known bug fix. This analysis also states if a reverse of the identified fix pattern can enable the bug to be reached and triggered. 71% of injection sites are dropped on the base of reachability. The authors could reintroduce over 8 000 bugs in 10 programs using just three bug patterns. The Forward-Porting approach in this work differs in that it reintroduces real-world vulnerabilities that, by definition, were present and exploitable in the program it is reintroduced in. Also, if the three patterns are generic, they rely on single-location statements like loops and `if` conditions. In our case, we manage fixes and breaking commits ranging across several files, functions and locations –that may include one (e.g., CVE-2020-24369's fix) or several of the three bug patterns– and their interaction with each other. Finally, the PoC-based approach in this work grants that all CVE reintroduced are reachable and exploitable, while the FixReverter fuzzing campaign could only reach 24% and exploit 8% of them. IntJect [37] introduces vulnerabilities in genuine code at the bytecode level for five distinct CWEs using Semantic Preserving Programs Transformation and Neural Machine Translation. Their method creates a synthetical benchmark by design as their model is trained on the Juliet test suite – a collection of artificial granular test cases, each representing a specific CWE [39].

*D. Software Evolution*

As early as 1980, it has been observed that a software change can fall into four different categories [40]. Either it is improving the code or correcting some part of it. It can also adapt to a new platform/new standards or be a preventative measure: a refactoring to ease future maintenance. These categories have since been extended [41]. This categorisation matters as monitoring and assessing code change similarities should enable developers to recognise software change patterns and elect the best way to deliver such changes [42].

Further, Lehman summarizes the laws of real-world software evolution [43]. The first law captures that software will always change with evolving requirements. These changes increase complexity unless work is done to reduce it (law II). Further, in absence of quality assessment (e.g., testing), software's quality decreases over its evolution (law VII). Our contribution (and the Magma benchmark principle) aims to improve law VII by improving fuzz-testing, while laws I and II imply a substantial effort to maintain a forward-ported groundtruth fuzzing benchmark.

In our forward-porting efforts, we encountered sequences of successive commits which can be linked to change bursts [44]. The authors showed that change bursts can be used to predict defective components in software. Thus, it seems likely that forward-porting a vulnerability to the latest software version will encounter many commits that hinder portability.

## VI. CONCLUSION

In this work, we reassessed the vulnerability *forward-porting* process used by Magma four years after its release. From the results, it became evident that a trivial *forward-porting*, as was used by Magma, is not sustainable. For 15 CVEs (47%) the software evolution has modified the source code to an extent, where reversing just the initial vulnerability fixing patch no longer suffices. Yet, this does not invalidate the general notion of *forward-porting*. We used a dichotomy to find the commit breaking the trivial *forward-porting* and analysed the code changes. In more than half of the failing cases (53%) reintroducing the vulnerability beyond the breaking commit requires only little code modification, such as renaming or changing variable types, or reverting upstream input checks. Our study results in 26 cases (76%) for which the amount of changes necessary is applicable to manual maintenance of the *forward-portability* of the benchmark. For the remaining CVEs, we carry the process to limits of manual maintenance, draw lessons and list the challenges that build toward the automation of the *forward-porting* process.

We further establish rules to maintain a qualitative *forward-ported* based benchmark over time based on incompatibility and functionality, above the complexity mentioned above. Our future research will focus on building such an automation (1) overcoming the challenges of extensive commits and expansive list of breaking commits, and (2) evaluating the *forward-porting* mechanism on a larger dataset of CVEs.